\documentstyle[preprint,aps]{revtex}
\begin{document}
\draft

\title{Chern-Simons Soliton Dynamics in Modular Parameter Space
\thanks {This work is supported in part by funds provided by the U.S.
Department of Energy (D.O.E.) under contract
\#DE-AC02-76ER03069.}}

\author{Q. Liu\\
Center for Theoretical Physics \\
Laboratory for Nuclear Science \\
and Department of Physics \\
Massachusetts Institute of Technology \\
Cambridge, Massachusetts 02139}

\date{\today}

\maketitle

\begin{center}
Submitted to {\it Physics Letters B}
\end{center}

\begin{abstract}
We study dynamics of non-relativistic Chern-Simons solitons, both in
the absence and in the presence of external fields. We find that a
phase, related to the $1$-cocyle of the Galileo group, must be
included to give the correct dynamical behavior. We show that
interactions among Chern-Simons solitons are mediated by an effective
Chern-Simons gauge field induced by the solitons. In the two soliton
case, we evaluate analytically the effective interaction Lagrangian,
which previously was found numerically.
\end{abstract}

\vspace{\fill}
\noindent CTP \# 2248
\newpage

Recently, there has been a lot of study on Chern-Simons solitons \cite{1}.
Topological and non-topological solitons are found in a relativistic theory
\cite{2,3}, and non-topological solitons exist in a non-relativistic model
\cite{4}. Since Chern-Simons dynamics is closely related to the quantum Hall
effect and perhaps to high-$T_c$ superconductivity, it is very interesting to
study these solitons \cite{5,6}. Here, we analyze the dynamics of
non-relativistic Chern-Simons solitons in modular parameter space.

Manton originally proposed a method for studying soliton dynamics with specific
application to monopole scattering \cite{7}. His idea can be summarized as
follows: if there are no forces between static solitons, then at low energies,
the dynamics of the full field theory can be described approximately on a
finite dimensional space, where degrees of freedom are the modular
parameters of the general static solution. This method has been widely applied
to many other systems \cite{5,8,9,10,11}. In a recent application,
the statistical interaction among non-relativistic Chern-Simons solitons is
obtained \cite{6}.

However, since only first-order time derivatives appear in non-relativistic
Chern-Simons theory,
it is not clear that Manton's method is directly applicable. Here this
question is analyzed and we find that in order to find the correct dynamical
behavior of non-relativistic Chern-Simons solitons, a phase, related to the
$1$-
cocycle of the Galileo group, must be introduced
 when applying the collective coordinate method. Moreover, we use this
modified method to study interactions both among well-seperated solitons and
between these solitons and external fields.

Let us define the notation: we shall use superscripts
$m, n = 1,..., N$ as soliton indices, and subscripts $i, j = 1, 2$ as
space indices, for which a summation convention is employed.

We first introduce the Jackiw-Pi Lagragian \cite{1},
\begin{equation}
L=\int d^{2} {\bf r} ( {\kappa \over 4} \epsilon^{\alpha\beta\gamma}
 A_{\alpha} F_{\beta\gamma} + i \psi^{\ast} D_t \psi - {1 \over 2} | {\bf D}
 \psi |^{2} + {1 \over {2 \kappa}} ( \psi^{\ast} \psi )^{2} )
\label{eq:jp}
\end{equation}
where $D_t = \partial_t + i A_{0}$, ${\bf D} = \nabla - i {\bf A}$.

The static solution for N solitons is
\begin{equation}
\psi_s = \rho^{ 1 \over 2} e^{ i \omega }
\label{eq:so}
\end{equation}
with
\begin{equation}
\rho = { {4 \kappa | f^{\prime} |^{2} } \over { (1 + | f |^{2} )^{2} } },
{}~~~~~\omega = Arg(f^{\prime}V^{2})
\label{eq:rho}
\end{equation}
where
\begin{equation}
f(z) = \sum_{m=1}^{N} {c^{m} \over z - a^{m}}, ~~~~~
V(z) = \prod_{m=1}^{N} ( z - a^{m} )
\label{eq:f}
\end{equation}
here $z=x+iy$, $f^{\prime} = { df \over dz}$ and $\omega$ is defined in
(\ref{eq:rho}) to make the solution nonsingular. Physically, $a^{m}$ is the
position of the $m$th soliton, and $c^{m}$ parametrizes its size and
phase. Note, the static solution $\psi_s$ satisfies the self-dual equation
\cite{1},
\begin{equation}
(D_{1} - i D_{2})\psi_s = 0
\label{eq:sd}
\end{equation}
and the action $I =\int dt L$ can also be written as \cite{1}
\begin{equation}
I = \int dt d^{2} { \bf r} ( i \psi^{\ast} \partial_t \psi - { 1 \over 2} |
( D_1 - i D_2) \psi |^{2})
\label{eq:ac}
\end{equation}

Now we discuss the dynamics of these solitons. For sake of simplicity, we hold
the $c^{m}s$ time-independent, and let the $a^{m}s$ be time-dependent. First,
we
notice: because of the Galileo invariance of our model, the static one-soliton
solution acquires a phase $ \Theta= {\bf v} \cdot {\bf r} - {1 \over 2} v^2 t
$ when boosted with a constant velocity  ${\bf v}$ \cite{1}. Motivated by this
 fact, we consider the following function,
\begin{equation}
\psi = \psi_s e^{i \Theta}
\label{eq:tr}
\end{equation}
where $\psi_s$ is the self-dual solution (\ref{eq:so}) and $\Theta$ is a
funtion of $\dot{a}^{m}$, $a^{m}$, $t$ and ${\bf r}$.

We assume that the time-evolution of well-separated Chern-Simons solitons
at low energies is approximately described by the effective Lagrangian for the
$a^{m}s$, which is obtained by substituting (\ref{eq:tr}) into the original
action I (\ref{eq:ac}). Notice that $\psi_s$ continues to satisfy equation
(\ref{eq:sd}) even with time-dependent parameters, hence we obtain
\begin{equation}
I_{eff} = \int dt d^{2} {\bf r} ( - \rho \partial_t \Theta - \rho \partial_t
\omega
 + {i \over 2} \partial_t \rho - {1 \over 2} \rho \partial_i \Theta \partial_i
\Theta )
\label{eq:z}
\end{equation}
Since ${d \over dt} \int d^{2} {\bf r} \rho = 0$,
\begin{equation}
I_{eff} = \int dt d^{2} {\bf r} ( - \rho \partial_t \Theta - \rho \partial_t
\omega - {1 \over 2} \rho \partial_i  \Theta \partial_i \Theta )
\label{eq:y}
\end{equation}

In order to determine $\Theta$, we
shall require that near the center of each soliton $\psi$ satisfy the
equation of motion of the original Lagrangian to order $\dot{\bf a}$; we
also assume that $\ddot{\bf a}$ is much smaller than $\dot{\bf a}$. This leads
to,
\begin{equation}
\Theta = \sum_{m} (\dot{\bf a}^{m}(t) G^m(\bf r)) \cdot {\bf r}
\label{eq:one}
\end{equation}
and, $G^{m} ({\bf r}) \longrightarrow 1$, when ${\bf r}$ is near ${\bf a}^{m}$;
while $G^{m} ({\bf r})
\longrightarrow 0$, when ${\bf r}$ is far away from ${\bf a}^m$. Also the
derivatives of $G^{m}$ are order $ (\dot {\bf a})^2 $, hence can be set
to zero. Thus, the effctive action becomes,
\begin{equation}
I_{eff} = \int dt d^{2} {\bf r} (- \sum_{m} \rho \partial_t ( \dot{\bf a}^{m}
\cdot {\bf r} G^{m} ) - \rho \partial_t \omega
- {1 \over 2} \sum_{m} \rho (\dot{\bf a}^m G^m) \cdot (\dot{\bf a}^m G^m) )
\label{eq:x}
\end{equation}
or after an integration by parts,
\begin{equation}
I_{eff} = \int dt d^{2} {\bf r} ( \sum_{m} \partial_t \rho (\dot{\bf a}^{m}
G^m)
\cdot {\bf r}  - \rho \partial_t \omega - {1 \over 2} \sum_{m}
\rho (\dot{\bf a}^m G^m) \cdot (\dot{\bf a}^m G^m)  )
\label{eq:w}
\end{equation}
Here an end point contribution has been dropped. Thus, our effective Lagrangian
is,
\begin{equation}
L_{eff} = \int d^{2} {\bf r}( \sum_{m} \partial_t \rho (\dot{\bf a}^{m} G^m)
\cdot {\bf r}  - \rho \partial_t \omega -  {1 \over 2} \sum_{m} \rho
(\dot{\bf a}^m G^m) \cdot (\dot{\bf a}^m G^m)  )
\label{eq:v}
\end{equation}

We divide $L_{eff}$ into two parts, $L_1$ and $L_2$, in which
 $L_1$ is the part induced by the phase $\Theta$ and $L_2$ is the part
obtained by direct application of Manton's prescription.
\begin{eqnarray}
L_{eff} = L_1 + L_2 \nonumber \\
L_1 = \sum_m \int d^{2} {\bf r} (\partial_t \rho  (\dot{\bf a}^{m} G^m) \cdot
  {\bf r}   - {1 \over 2} \rho (\dot{\bf a}^m G^m) \cdot (\dot{\bf a}^m
G^m)) \nonumber \\
L_2 = - \int d^{2} {\bf r}  \rho \partial_t \omega
\label{eq:t}
\end{eqnarray}

We first evaluate $L_1$. Using the above described properties of $G^m$, we
obtain,
\begin{eqnarray}
L_1 = \sum_m {\dot{\bf a}^{m} \cdot \int d^{2} {\bf r} \partial_t (\rho G^m
{\bf r} ) - {1 \over 2} \dot{\bf a}^m  \cdot \dot{\bf a}^m
\int d^{2} {\bf r} \rho^m} \nonumber \\
    = \sum_m {\dot{\bf a}^{m} \cdot {d \over dt} (\int d^{2} {\bf r} \rho^m
{\bf r} ) - {1 \over 2} \dot{\bf a}^m  \cdot \dot{\bf a}^m
\int d^{2} {\bf r} \rho^m}
\end{eqnarray}
where $ \rho^m $ is the spherically symmetric one-soliton density for the $m$th
soliton.
Finally, using $\int d^{2} {\bf r} \rho^{m} {\bf r} = 4 \pi \kappa {\bf a}^m$
and $\int d^{2} {\bf r} \rho^{m} = 4 \pi \kappa$, leaves,
\begin{equation}
L_1 = 2 \pi \kappa \sum_{m} \dot{\bf a}^{m} \cdot \dot{\bf a}^{m}
\label{eq:ki}
\end{equation}
Not surprisingly, the familiar kinetic energy term for non-relativistic
particles is recovered and the mass $4 \pi \kappa $ is exactly what we expect
from a consideration of the single-soliton momentum \cite{1}.

Now we evaluate $L_2$. Notice that $f^{\prime} V^{2}$ can always be written as,
\begin{equation}
f^{\prime} V^{2} = -(\sum_{m = 1}^{N} c^m)   \prod_{n =1}^{2 N - 2} (b^{n} - z)
\end{equation}
where each $b^{n}$ solves the following equation,
\begin{equation}
\sum_{m=1}^{N} ( c^{m} \prod_{n \neq m, n=1}^{N} (z - a^{n})^{2}) = 0
\label{eq:st}
\end{equation}
Thus, we have,
\begin{equation}
\omega = Arg (f^{\prime} V^2) = \sum_{n=1}^{2N-2} Arg (b^{n} -z) + const
\end{equation}
Using the correspondence between a complex number $z$ and a real 2-dimensional
vector ${\bf r}$ as well as the formula $Arg(z) =  \theta({\bf r}) \equiv
tan^{- 1} ({r_2 \over r_1})$, we have,
\begin{eqnarray}
L_2 = - \sum_{n=1}^{2N - 2} \int d^{2} {\bf r} \rho \partial_t \theta
({\bf b}^{n} -{\bf r})  \nonumber \\
    =  - \sum_{n=1}^{2N - 2}  \dot{\bf b}^{n} \cdot \int d^{2} {\bf r}
{\partial \over \partial {\bf b}^{n}} \theta ({\bf b}^{n} -{\bf r}) \rho
\label{eq:ll}
\end{eqnarray}
Recall that in the original theory the Chern-Simons vector potential is given
 by \cite{1},
\begin{equation}
{\bf A} ({\bf r}, t ) = - { 1 \over 2 \pi \kappa } \int d^{2} {\bf r}^{\prime}
 \nabla \theta ({\bf r} -{\bf r}^{\prime}) \rho ({\bf r}^{\prime},t)
\end{equation}
and we see that (\ref{eq:ll}) becomes,
\begin{equation}
L_2 = 2 \pi \kappa \sum_{n=1}^{2 N - 2}  \dot{\bf b}^{n} \cdot {\bf A}
({\bf b}^{n}, t)
\label{eq:int}
\end{equation}
The interaction among the non-relativistic Chern-Simons solitons is mediated
by an effective Chern-Simons vector potential induced by these solitons. A
similar result was obtained in the relativistic Chern-Simons model by Kim and
Min \cite{5}.

We can further simplify (\ref{eq:int}). As shown in Ref. \cite{1}, for the
self-dual solution,
\begin{equation}
{\bf A} = -{1 \over 2} \nabla \times ln \rho + \nabla \omega
\end{equation}
Defining $\Phi ( {\bf r} ) = ( | V |^{2} + | f V |^{2})$, we get,
\begin{equation}
L_2 = 2 \pi \kappa \sum_{n=1}^{2N - 2} \dot{\bf b}^{n} \cdot ( \nabla \times
ln \Phi ( {\bf r} ))|_{{\bf r} = {\bf b}^{n}}
\label{eq:po}
\end{equation}

Combining (\ref{eq:ki}) with (\ref{eq:po}), we have,
\begin{equation}
L_{eff} = 2 \pi \kappa \sum_{m} \dot{\bf a}^{m} \cdot \dot{\bf a}^{m} + 2
 \pi \kappa \sum_{n=1}^{2N - 2} \dot{\bf b}^{n} \cdot ( \nabla \times
ln \Phi ( {\bf r} ))|_{{\bf r} = {\bf b}^{n}}
\end{equation}
where ${\bf b}^{n}$ is determined by solving (\ref{eq:st}).

In principle, the interaction term $L_2$ for a fixed $N$ can be simplified
with the aid of (\ref{eq:st}). As an example, we shall make the simplification
 for two solitons in their center of mass frame,
${\bf a}^1 = - {\bf a}^2 = {\bf a}$, and arbitary constants $c^1$ and $c^2$.
In this case, equation (\ref{eq:st}) becomes,
\begin{equation}
( c^{1} + c^{2} ) ( z^{2} + 2 d z a + a^{2} ) = 0
\label{eq:ao}
\end{equation}
where $ d = {c^{1} - c^{2} \over c^{1} + c^{2} }$.

Equation (\ref{eq:ao}) has two roots $b^{1,2} = a (-d \pm  \sqrt{d^2-1} )$
while $\dot{b}^{1,2} = \dot{a} (-d \pm \sqrt{d^2-1} )$.
Substituting these into the effective Lagrangian and using
the explict form of $\Phi$ for two solitons, we obtain, in complex notation,
\begin{eqnarray}
L_2 =  -4 \pi \kappa \sum_{n=1}^{2} Im( \dot{b}^n \partial_z ln \Phi(z))
|_{z=b^n}  \nonumber \\
    = - 4 \pi \kappa \sum_{n=1}^{2} Im(  {\dot{a}^n a^{\ast} \over |a|^2}
({(-1)^n d ({\sqrt{d^2-1}})^{\ast} \over | \sqrt{d^2-1} |^2} + 1))
\nonumber \\
    = - 8 \pi \kappa Im ( {\dot{a}^n a^{\ast} \over |a|^2} ) \nonumber \\
    = - 8 \pi \kappa {d \over dt} Arg(a) = - 8 \pi \kappa {d \over dt}
\theta({\bf a})
\label{eq:ase}
\end{eqnarray}
Recalling that $\theta ( {\bf a} )$ is the relative angle between two solitons,
we see that (\ref{eq:ase}) is the statistical interaction with spin $S = - 4
\pi \kappa$ \cite{12}. This coincides with the result obtained by Hua and
Chou with numerical integration \cite{6}. Thus, classically, our Lagrangian
describes two free-moving non-relativistic particles with statistical
interaction.

As another interesting example, we apply our method to
a single Chern-Simons soliton in the presence of external fields. To be
specfic, we
consider one soliton either in a constant external electric field or in a
constant external magnetic field or in a harmonic potential or
in any combination of these three. The most general action is
\begin{equation}
I = \int dt d^{2} { \bf r} ( i \psi^{\ast} \partial_t \psi - A^e_0 \psi^{\ast}
 \psi - { 1 \over 2} k r^2 \psi^{\ast} \psi - { 1 \over 2} |
( D_1^e - i D_2^e) \psi |^{2})
\label{eq:newac}
\end{equation}
where ${\bf D}^e = \nabla - i {\bf A} -i {\bf A}^e$, $ A^e_i = - { 1 \over 2 }
 \epsilon_{ij} r_j B$, $A^e_0 = - { \bf r} \cdot { \bf E}$ and $k$ is the
strength of the harmonic potential .

In this case, we choose the phase $\Theta$ as follows,
\begin{equation}
\Theta = \dot{\bf a} \cdot {\bf r} - {1 \over 8} B^2 \int^t d t^{\prime}
{\bf a}(t^{\prime}) \cdot {\bf a}(t^{\prime})
\end{equation}
Then we substitute the trial function (\ref{eq:tr}) with this new $\Theta$
 into (\ref{eq:newac}), following similar procedures, we obtain,
\begin{equation}
L_{eff} = {1 \over 2} \dot{\bf a} \cdot \dot{\bf a} \int d^2 {\bf r} \rho +
\int d^2 {\bf r} \rho {\bf E} \cdot {\bf r} - { 1 \over 2} k \int d^2 {\bf r}
\rho r^2 +  \int d^2 {\bf r} \rho \dot{\bf a} \cdot {\bf A}^e({\bf r})
\end{equation}
By using the spherical symmetry of the one-soliton solution and the explicit
form of ${\bf A}^e$, we have,
\begin{equation}
L_{eff} = 2 \pi \kappa \dot{\bf a} \cdot \dot{\bf a} + 4 \pi \kappa
{\bf a} \cdot {\bf E} - 4 \pi \kappa \dot{\bf a} \cdot
{\bf A}^e({\bf a}) - 2 \pi \kappa k |\bf a|^2
\end{equation}
Here an irrelevant constant term is dropped.

Thus, we see that the Chern-Simons soliton behaves like a non-relativistic
point-like particle with charge $4 \pi \kappa$ and mass $4 \pi \kappa$ in
these external fields. In fact, exact soliton solutions in the presence of
these
fields can be found by a coordinate transformation \cite{13}. Our result agrees
with the behavior of these solutions.

In summary, we comment on the phase $\Theta$. From our work,
it is  clear that this phase plays a very important role in the dynamics of
non-relativistic Chern-Simons solitons. However, we have not
determined this phase exactly. Thus, it would be very interesting to look for
some method to find this phase.

The author thanks Prof. R. Jackiw for suggesting this problem and many helpful
 discussions. He also thanks D. Bak, Dr. C. Chou and Dr. C. Lee for
valuable discussions and comments.

\end{document}